\documentclass[twocolumn,showpacs,prb]{revtex4}
\usepackage{graphicx}

\begin{document}
\title{Theory of a two-level artificial molecule in
laterally coupled  quantum Hall droplets}

\author{Ramin M. Abolfath, W. Dybalski, and Pawel Hawrylak}

\affiliation{
Institute for Microstructural Sciences,
National Research Council of Canada,
Ottawa, K1A 0R6, Canada}

\date{\today}

\begin{abstract}
We present a theory of laterally coupled quantum Hall droplets with 
electron numbers (N1,N2) at filling factor $\nu=2$. We show that the 
edge states of each droplet are tunnel coupled and form
a two-level artificial molecule. By populating the edge states with
one electron each a two electron molecule is formed. We predict the
singlet-triplet transitions of the effective two-electron molecule as a function
of the magnetic field, the number of electrons, and confining potential using
the configuration interaction method (CI) coupled with 
the unrestricted Hartree-Fock (URHF) basis. In addition to the 
 singlet-triplet transitions
of a 2 electron molecule involving edge states, triplet 
transitions involving transfer
of electrons to the center of individual dots exist for 
$(N1 \geq 5, N2 \geq 5)$. 
\end{abstract}

\pacs{73.43.Lp}

\maketitle

\section{Introduction}

\begin{figure}
\begin{center}
\vspace{1cm}
\includegraphics[width=0.7\linewidth]{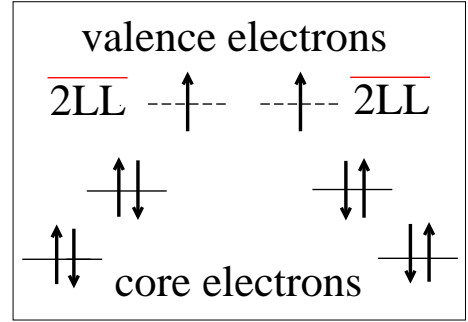}
\caption{$\nu=2$ droplets}
\label{EdgeOrbitals}
\end{center}
\end{figure}

There is currently significant experimental 
\cite{Livermore,OOSTERKAMP,Holleitner,ciorga,kouwenhoven,MichelPRL93,petta,
Rontani,Petta,Craig,Hatano,Pioro-Ladriere} 
and theoretical 
\cite{palacios,brum,koskinen97,loss-divincenzo,burkard-loss,
Wensauer00,HuDasSarma,Yannouleas,HarjuPRL02,WKH,Wojtek} 
interest in coupled lateral quantum dots. 
The main effort is 
on developing means of control of quantum mechanical coupling of 
artificial molecules, 
with each dot playing the role of an artificial 
atom\cite{palacios,kouwenhoven,MichelPRL93,petta}.
This is stimulated
largely by the prediction that with one electron in each dot, 
the magnetic field driven singlet-triplet spin transitions
of a two-level molecule have potential application as quantum 
gates \cite{brum,loss-divincenzo,burkard-loss,HuDasSarma}. 
Recent experiments by Pioro-Ladriere et al. in Ref.\onlinecite{MichelPRL93}
suggested that one can create an effective two-level molecule in many electron
lateral quantum dots by the application of high magnetic field.
In strong magnetic field electrons are expected to form quantum Hall droplet in each
quantum dot. Edge states of each droplet can be coupled  
in a controlled way using barrier electrodes, and at filling factor $\nu=2$  
effectively reduce the many-electron-double dot system to a 
two-level molecule \cite{MichelPRL93}. 
When populated with one electron each, 
one expects to have singlet triplet transitions of two valence electrons in the
background of core electrons of the spin singlet $\nu=2$ droplets. 
This is schematically  illustrated in 
Fig.\ref{EdgeOrbitals} where  lateral lines correspond to 
energies of orbitals of the lowest 
Landau level in each droplet as a function of position,
 with solid lines being occupied by the spin up/down
core electrons, and dashed lines by the valence electrons in edge states. 
In analogy with a single dot \cite{ciorga,WKH} the orbitals of the second Landau level
of each dot  are also shown. The energy of these  orbitals becomes comparable to 
the energy
of valence electrons when the number of electrons is $\geq 5$ per dot. For these
electron numbers it is possible in single dots to move electrons from the 
edge valence orbitals 
to 2LL orbitals localized in the center. 

These qualitative considerations are supported by a presented here 
microscopic  theory of 
laterally coupled  quantum Hall droplets. 
The quantum dot molecules studied in this work differ from the single
dots by the complex confining potential, and by
twice as many electrons.
We address these difficulties by extending
the combination of the Hartree-Fock and exact diagonalization
configuration interaction techniques, introduced to study single
dots in Ref. \onlinecite{WKH}, to real space calculations
for large electron numbers in double dots.
This allows us to predict the magnetic field induced singlet-triplet spin transitions
of quantum Hall droplets with $(N1+1,N2+1)$ electron numbers,
interpret them in terms of an effective two-electron (1,1) artificial molecule,
and determine
exchange coupling constant J as a function of the number of core electrons and
the magnetic field. The possibility of storing electrons in 2LL orbitals 
is demonstrated.

\section{Hamiltonian}

Our theory is based on effective mass envelope function to describe the
confined electrons in quantum dots. We consider electron motion to be
quasi-two-dimensional and coupled to the perpendicular external magnetic 
field  by vector potential $A$. 
With the total number of electrons $N=N1+N2$
the quantum dot molecule Hamiltonian can be written as:
\begin{equation}
H = \sum_{i=1}^N  T_i
+ \frac{e^2}{2\epsilon}\sum_{i \neq j}\frac{1}{|\vec{r}_i - \vec{r}_j|},
\end{equation}
where $T=\frac{1}{2m^*}\left(\frac{\hbar}{i}\nabla 
+ \frac{e}{c} A(\vec{r})\right)^2 + V(x,y)$ is the 
one electron Hamiltonian with $V(\vec{r})$ the 
quantum dot molecule confining potential,
$m^*$ the conduction-electron effective mass, 
$e$ the electron chrage, and
$\epsilon$ the host semiconductor dielectric constant. 
The Zeeman spin splitting (very small for GaAs) is neglected here.
In what follows we use GaAs effective atomic energy and length 
units with  $Ry^*=5.93 meV$, and $a^*_0 = 9.79 nm$.
The choice of gauge $A$ plays significant role in
improving the numerical accuracy of single particle spectrum. 
Because the quantum dot molecules considered in 
this study are weakly coupled, we adopt
a gauge field to separate the vector potential $A$ into two
parts: $A=A_L$ if $x<0$ and $A=A_R$ otherwise.
To distinguish the electrons localized in the different dots 
we use the pseudo-spin labels left (L) and right (R).
Here $A_L=(B/2)(-y,x+a)$, and $A_R=(B/2)(-y,x-a)$ are
localized vector potentials at the center of each dot,
$B$ is the magnetic field, and $2a$ is the inter-dot
separation.
The associated wavefunctions for electron in the left and
right dot are connected via a gauge transformation
$\psi_L(\vec{r}) =  \psi_R(\vec{r})
\exp\left(-i\hbar\omega_c\frac{ay}{2}\right)$
\cite{ramin pawel}.

\section{Confining potential}

The double quantum dot potential $V(x,y)$ defined by electrostatic 
gates is characterized by
two potential minima. With our focus on electronic correlations, 
we parameterize electrostatic potential of a general class of coupled
 quantum dots by a sum of three Gaussians  \cite{HuDasSarma} 
$V(x,y)=V_L~ \exp[{-\frac{(x+a)^2+y^2}{\Delta^2}}]
        +V_R~ \exp[{-\frac{(x-a)^2+y^2}{\Delta^2}}]
+V_p \exp[{-\frac{x^2}{\Delta_{Px}^2}-\frac{y^2}{\Delta_{Py}^2}}]$.
Here $V_L,V_R$ describe the depth of the left and right quantum dot minima
located at $x=-a,y=0$ and $x=+a,y=0$, and $V_p$ is the plunger gate potential
controlled by the central gate. For identical dots, $V_L=V_R=V_0$, 
and confining potential  exhibits inversion symmetry. In the following 
numerical examples we will 
parameterize it by $V_0=-10, a=2, \Delta=2.5$, 
and $\Delta_{Px}=0.3$, 
$\Delta_{Py}=2.5$, in effective atomic units. 
$V_p$, which controls the potential barrier,
is varied between zero and $10 Ry^*$, independent 
of the locations of the quantum dots.
The choice of parameters ensures weakly coupled quantum dots.

\section{single particle spectrum}

The potential of each isolated dot is a single Gaussian potential. Expanding
it in the vicinity of the minimum yields a
parabolic potential $V(r)=m^* \omega_0^2 r^2/2$ with the strength 
$\omega_0=2\sqrt{|V_0|/\Delta^2}$.  
The low energy spectrum of each dot corresponds to two
harmonic oscillators with eigen-energies
$\varepsilon_{nm}=\hbar\omega_+(n+1/2)+\hbar\omega_-(m+1/2)$.
Here  $\omega_\pm = \sqrt{\omega_0^2 + \omega_c^2/4} \pm \omega_c/2$, $\omega_c$
is the cyclotron energy, and
$n,m=0,1,2,...$.
With increasing magnetic field the $\hbar\omega_-$ decreases to zero
while $\hbar\omega_+$ approaches the cyclotron energy $\omega_c$, and
the states $|m,n\rangle$ evolve into the nth Landau level.
The $\nu=2$ spin singlet quantum Hall droplet is formed 
if 2N electrons occupy the 
N successive $|m,n=0 \rangle$ lowest Landau level (LLL) orbitals. 
When  extra  $2N+1$th electron
is added it  occupies the edge orbital $|m=N,n=0 \rangle$. 
These $(2\times N+1,2 \times N+1)$ configurations, for the
two isolated dots, are shown in Fig.\ref{EdgeOrbitals}. 
In each isolated dot increasing the
magnetic field leads to spin flips while 
decreasing the magnetic field lowers the energy of the edge 
$|m=N,n=0 \rangle$ orbital
with respect to the lowest unoccupied center orbital $|m=0,n=1 \rangle$ of the 
second Landau level (2LL). 
At a critical magnetic field a LL crossing occurs 
and the electron transfers from the edge orbital to central orbital,
leading to redistribution of electrons from the edge to center.\cite{ciorga,WKH}

\begin{figure}
\begin{center}
\vspace{1cm}
\includegraphics[width=0.98\linewidth]{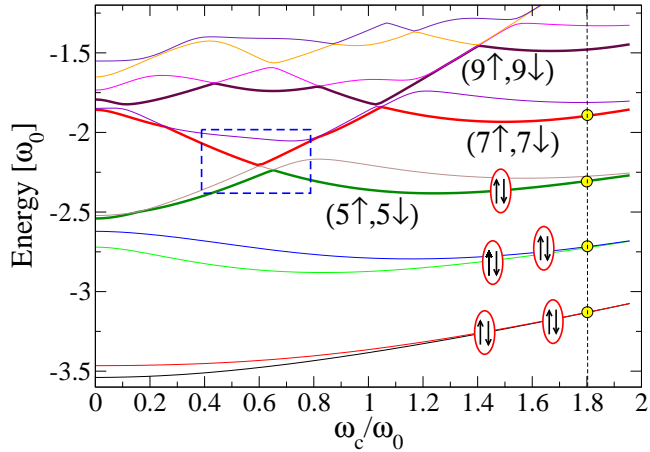}
\caption{Double dot single particle spectrum vs. cyclotron energy.
Molecular states can be grouped by their parity 
(or pseudo-spin quantum number $\pm$).
States with $j=$odd (even) have positive (negative) parity (pseudo-spin). 
States with opposite parity cross
and states with the same parity anticross.
An enlarged part of spectrum, inside the box surrounded by the dashed lines,
is shown in Fig. \ref{wavefunctions5}.
}
\label{E_sp}
\end{center}
\end{figure}

We now turn to the description of coupled dots in strong magnetic field
($n=0$).
As a first approximation one might expect the $|m; L \rangle$ orbitals of the left dot $L$
to be coupled only with the corresponding $|m; R \rangle$ orbitals of the right (R) dot,
and form a pair of the symmetric and anti-symmetric orbitals 
\begin{equation}
|m,\pm\rangle=\frac{1}{\sqrt{2(1\pm S_m)}}(|m,L\rangle\pm|m,R\rangle),
\end{equation}
where $S_m=Re(\langle m,L|m,R\rangle)$.
Therefore in high magnetic field we expect the formation of
shells of closely spaced pairs of bonding-antibonding levels.
This is illustrated
in  Fig. \ref{E_sp} which shows the magnetic field evolution of 
the numerically calculated single particle spectrum of a  double dot.
For the illustration we scale the unit of energy by 
$\hbar\omega_0\approx 2.53 Ry^*$,
using the parameters of the confinement potential
with the energy barrier set to $V_p=7Ry^*$.
The single particle eigenvalues and eigenvectors calculated by
diagonalizing $T$ are given 
by $\tilde{\epsilon}_j$ and $\tilde{\varphi}_j$ respectively.
The spectrum is calculated accurately by discretizing in real space 
the single particle Hamiltonian $T$ using special gauge transformation. 
The resulting large matrices are diagonalized  
using conjugate gradient algorithms \cite{ramin pawel}.

At zero magnetic field Fig. \ref{E_sp} shows the formation of 
hybridized S, P, and D shells, to be discussed elsewhere \cite{ramin pawel}.
In high magnetic field the pairs of closely spaced
levels $ |m,\pm \rangle $ separated by $\approx \omega_-$ are clearly visible. 
The energy spacing between the levels in each pair increases
at higher energy part of spectrum, 
because the potential barrier is less effective
and tunneling becomes stronger.

\begin{figure}
\begin{center}
\vspace{1cm}
\includegraphics[angle=-90,width=1.0\linewidth]{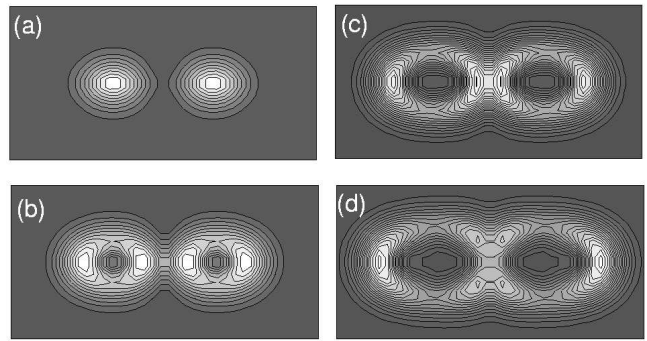}
\caption{Single electron densities at $\omega_c=1.8 \omega_0$ 
correspoponding to $j=1,3,5$, and $j=7$ (from a to d).
}
\label{WF48}
\end{center}
\end{figure}

The existence of pairs of levels in the numerically obtained 
spectrum suggests that the corresponding wavefunction 
$|\tilde{\varphi}_j\rangle$
does indeed admit a description in terms of orbitals localized
in each dot. This is confirmed by
the calculated overlaps $\langle\tilde{\varphi}_j|m,\pm\rangle$
shown in Table \ref{etable} as well as by    
the corresponding probability densities  shown in Fig.\ref{WF48}.

\begin{table}
{\centering
\begin{tabular}{|c|c|c|}
\hline
$\omega_c=1.8\omega_0$ & $\tilde\epsilon_j [\omega_0]$  
& $\epsilon_{LCAO} [\omega_0]$\\ \hline
$|\langle 1|1+\rangle|^2=0.99$     &  -3.13   & -3.12 \\ \hline 
$|\langle 2|1-\rangle|^2=0.99$     &  -3.12   & -3.12 \\ \hline 
$|\langle 3|2+\rangle|^2=0.98$     &  -2.72   & -2.71 \\ \hline 
$|\langle 4|2-\rangle|^2=0.99$     &  -2.71   & -2.70 \\ \hline 
$|\langle 5|3+\rangle|^2=0.95$     &  -2.30   & -2.28 \\ \hline 
$|\langle 6|3-\rangle|^2=0.99$     &  -2.27   & -2.26 \\  

\hline
\end{tabular}\vspace{1mm}\\}

\caption[]{
Comparison between real space molecular wave functions 
$|\tilde{\varphi}_j\rangle$ and the 
linear combination atomic orbitals (LCAO) $|m\pm\rangle$ is given by
the overlap $|\langle\tilde{\varphi}_j|m\pm\rangle|^2$
at $\omega_c=1.8\omega_0$.
The corresponding energies are shown.
\label{etable}}
\end{table}

\begin{figure}
\begin{center}\vspace{1cm}
\includegraphics[width=0.4\textwidth]{E_sp_62x31_Vp7_v8_omega0.eps}
\includegraphics[angle=-90,width=0.4\textwidth]{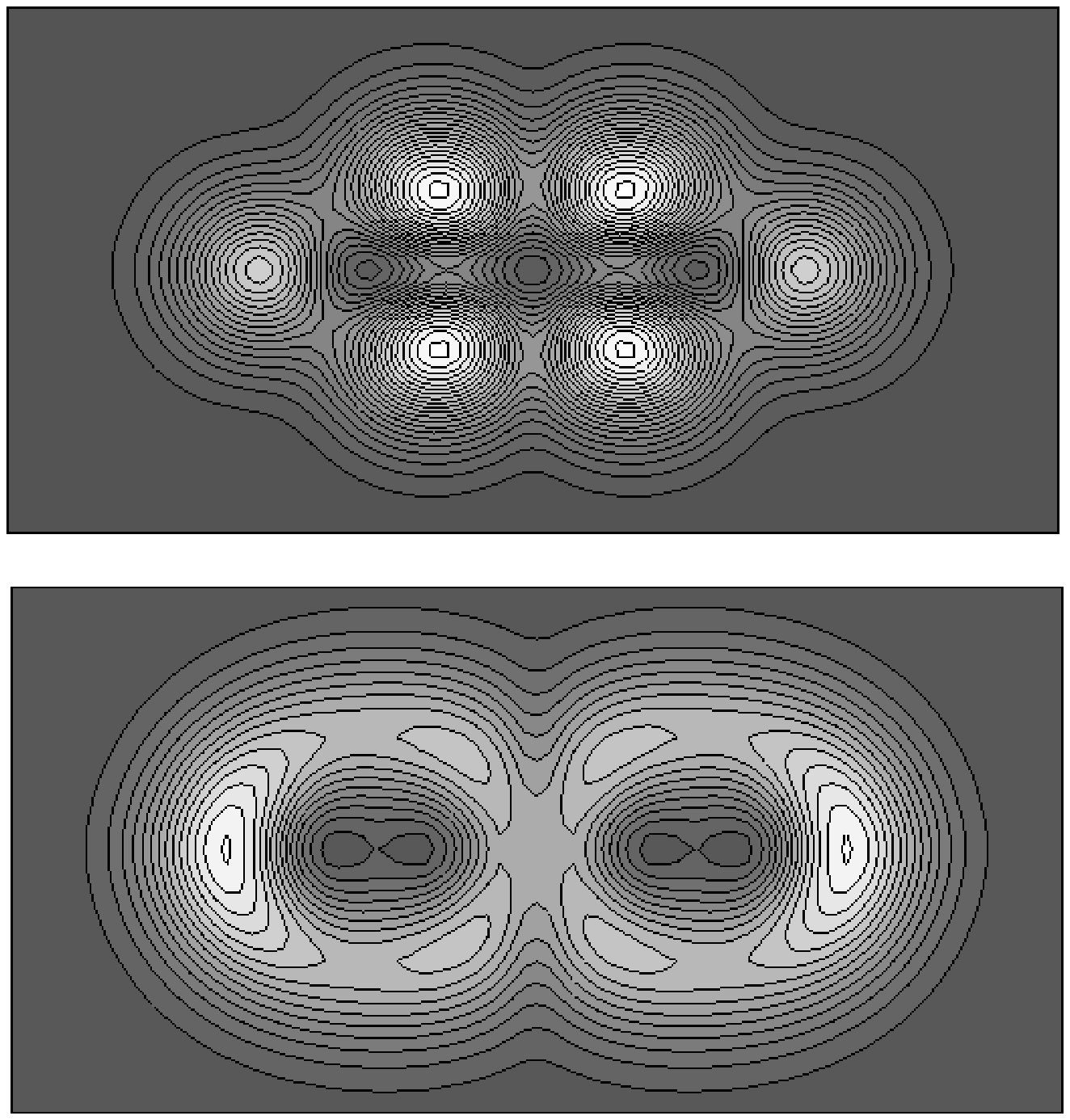}
\caption{
Top: Energies of single particle states with their parities are 
shown. 
States with positive and negative parities are represented by
pseudo-spin indices $|\pm\rangle$.
Bottom: The 5th single-particle orbital between two sides of Landau 
level crossing
is shown at $\omega_c=0.5\omega_0$, and  $\omega_c=0.8\omega_0$.
The top and bottom correspond to the innermost orbital
of second Landau level and outermost of the first Landau level.
}
\label{wavefunctions5}
\end{center}
\end{figure}

An important part of the  single particle spectrum is the $\nu=2$ phase
where electrons with opposite spin occupy the first Landau level orbitals
(in weak Zeeman coupling limit).
In a Fock-Darwin picture of a quantum dot spectrum, this corresponds to
the $|m,n=0\rangle$ orbitals
just after the last orbital crossing with the $|m=0,n=1\rangle$ orbital.
In coupled quantum dots, we are interested in transitions between the
states derived from the lowest Landau level and the states derived from the
second Landau level. The crossing of the two levels in a single dot
can occur for the 3rd and higher levels. A similar effect takes place 
in a coupled quantum dot but for the fifth and higher levels. 
The single-electron probability density 
of the 5th level at $\omega_c=0.5\omega_0$ 
and at $\omega_c=0.8\omega_0$ is shown in Fig. \ref{wavefunctions5}.
The electron which occupies the
5th level is closer to the center of each dot at
low magnetic fields.
It then moves to an edge configuration at higher fields.

From the single particle spectrum presented in Fig. \ref{E_sp}, 
the first set of electronic configurations which probe such crossing 
consists of 9-12 electrons occupying five lowest single particle states. 
The corresponding (5,5) half-filled shell configuration
is shown in Fig.\ref{E_sp}. The characteristic shell structure 
of a coupled dot in high magnetic field
leads to magnetic electron numbers 
populating different  $ |m,\pm \rangle $ electronic shells.  
The half-filled shells correspond to electron numbers $(N1,N2)=(1,1),
(3,3),(5,5),(7,7),\dots$ 
while filled shells correspond to $(N1,N2)=(2,2),(4,4),(6,6),(8,8),\dots$ 
configurations. We expect the half and fully filled shells to 
have special electronic properties.
In particular, the half filled shells offer the possibility of singlet-triplet
transitions. For the shells (5,5) and up  we expect to be able
to move  the valence electrons from the edge orbitals to the center orbitals.
The crossing of the edge and center orbitals is visible as a cusp in the energy of the
fifth molecular orbital shown as a bold line in Fig.\ref{E_sp}.

\section{Many body spectrum}
We now turn to the effect of electron-electron interactions.
Denoting the creation (annihilation) operators for electrons in
non-interacting SP state $|\alpha\sigma\rangle$ by 
$\tilde{c}^\dagger_{\alpha\sigma}~(\tilde{c}_{\alpha\sigma})$,
the Hamiltonian of an interacting system in second quantization
can be written as
\begin{eqnarray}
H&=&\sum_{\alpha\beta}\sum_{\sigma\sigma'} 
\langle \alpha\sigma | T | \beta \sigma' \rangle 
\tilde{c}^\dagger_{\alpha\sigma} \tilde{c}_{\beta\sigma'} 
\nonumber \\ &&
+ \frac{1}{2}\sum_{\alpha\beta\gamma\mu}\sum_{\sigma\sigma'}
\tilde{V}_{\alpha\sigma,\beta\sigma',\gamma\sigma',\mu\sigma}
\tilde{c}^\dagger_{\alpha\sigma} \tilde{c}^\dagger_{\beta\sigma'} 
\tilde{c}_{\gamma\sigma'} \tilde{c}_{\mu\sigma}, 
\label{multiparticle}
\end{eqnarray}       
where 
$\langle \alpha\sigma | T | \beta \sigma' \rangle
=\tilde{\epsilon}_{\alpha\sigma}\delta_{\alpha\beta}\delta_{\sigma\sigma'}$
and
$
\tilde{V}_{\alpha\sigma,\beta\sigma',\mu\sigma',\nu\sigma,} = 
\int d\vec{r} \int d\vec{r'}  
\tilde{\varphi}^*_{\alpha\sigma}(\vec{r})
\tilde{\varphi}^*_{\beta\sigma'}(\vec{r'}) 
\frac{e^2}{\epsilon |\vec{r}-\vec{r'}|}
\tilde{\varphi}_{\mu\sigma'}(\vec{r'})
\tilde{\varphi}_{\nu\sigma}(\vec{r})
$, are the single particle and
the two-body Coulomb matrix elements.
Here $\alpha,\beta,\mu,\nu$ are spatial orbital indices, and
$\sigma,\sigma'$ are the spin indices.
$\tilde{\epsilon}_{\alpha\sigma}$ is the single particle energy,
$e$ is the electric charge.

\subsection{Unrestricted Hartree-Fock Approximation (URHFA)}

\begin{figure}
\begin{center}
\includegraphics[width=0.9\linewidth]{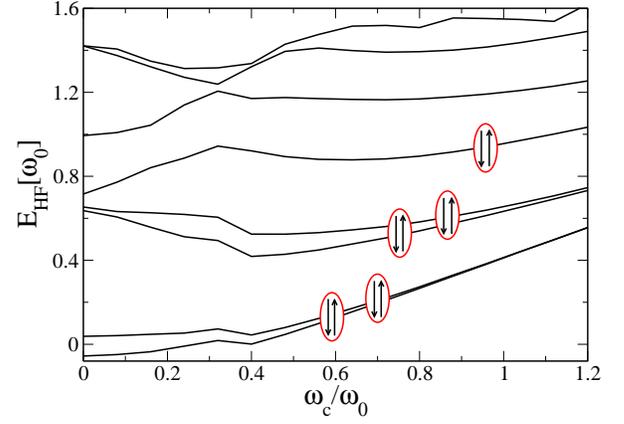}\vspace{1cm}
\noindent
\caption{
URHF eigen-energies vs. cyclotron energy  for 10 electrons.
The HF energy gap between HOMO (highest occupied molecular orbital),
and LUMO (lowest unoccupied molecular orbital) is clearly visible.
}
\label{E_HFvsB}
\end{center}\vspace{0.5cm}
\end{figure}

\begin{figure}
\begin{center}
\includegraphics[width=0.98\linewidth]{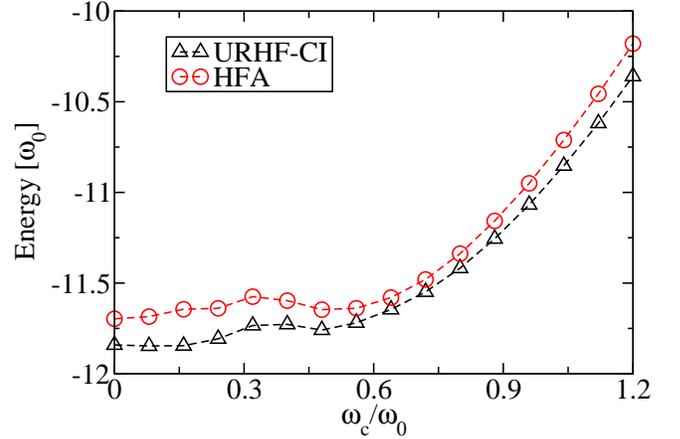}
\noindent
\caption{
URHFA, and URHFA-CI ground state energy vs. cyclotron energy are 
shown for a quantum dot molecule.
}
\label{URHF_10e}
\end{center}
\end{figure}

We now proceed to include electron-electron interactions in two steps:
direct and exchange interaction using unrestricted Hartree-Fock approximation
(URHF), and correlations using URHF basis in the configuration interaction method (URHF-CI). 
The spin-dependent HF orbitals $|\varphi_{i\sigma}\rangle $ are obtained 
from  the $N_l$ non-interacting
single particle orbitals $|\tilde{\varphi}_{\alpha}\rangle$,
 energy spectrum of which
is shown in Fig.\ref{E_sp},
 by the transformation
$|\varphi_{i\sigma}\rangle = \sum_{\alpha=1}^{N_l} a^{(i)}_{\alpha\sigma} 
|\tilde{\varphi}_{\alpha}, \sigma\rangle$.
The variational parameters $a^{(i)}_{\alpha\sigma}$
are solutions of self-consistent Pople-Nesbet 
equations \cite{Szabo_book}:
\begin{eqnarray}
&&\sum_{\gamma=1}^{N_l}
\{\tilde{\epsilon}_\mu\delta_{\gamma\mu}+
\sum_{\alpha,\beta=1}^{N_l}\tilde{V}_{\mu\alpha\beta\gamma}
[\sum_{j=1}^{N_\uparrow}
a^{*(j)}_{\alpha\uparrow} a_{\beta\uparrow}^{(j)} 
+ \sum_{j=1}^{N_\downarrow} \nonumber \\ &&
a^{*(j)}_{\alpha\downarrow} a_{\beta\downarrow}^{(j)} ] 
-\tilde{V}_{\mu\alpha\gamma\beta}
\sum_{j=1}^{N_\uparrow}
a^{*(j)}_{\alpha\uparrow} a_{\beta\uparrow}^{(j)}  
\} a_{\gamma\uparrow}^{(i)} 
= \epsilon_{i\uparrow} ~ a_{\mu\uparrow}^{(i)},
\label{urhfeq1}
\end{eqnarray}
where $\tilde{V}_{\alpha\beta\mu\nu}$ are Coulomb matrix elements
calculated using non-interacting single particle states.
A similar equation holds for spin down electrons.
The calculations are carried out for
all possible total spin $S_z$ 
configurations.
The calculated HF eigen-energies and 
total energies for the $N=10$ electrons with $S_z=0$  in a
magnetic field are shown in Figs. \ref{E_HFvsB} and \ref{URHF_10e}.
Comparing HF spectrum (Fig. \ref{E_HFvsB}) with the single particle 
spectrum (Fig. \ref{E_sp}),
one observes that a HF gap developed at the Fermi level,
between the highest occupied molecular state, and the lowest
unoccupied molecular state,
and the Landau level crossing 
between single particle Landau levels
has been shifted to lower magnetic fields,
from $\omega_c=0.6\omega_0$ in single
particle spectrum to $\omega_c=0.3\omega_0$ 
in HF spectrum.

\subsection{URHF configuration interaction method}

\begin{figure}
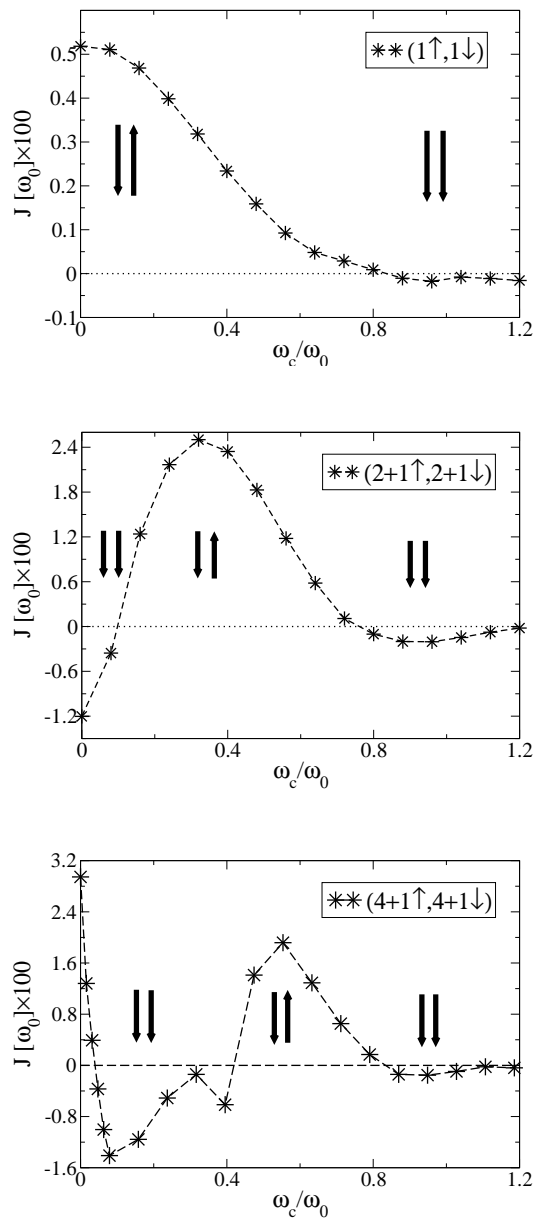

\begin{center}\vspace{-1cm}
\includegraphics[width=0.8\linewidth]{J2_omega0.eps}\vspace{0.8cm}
\includegraphics[width=0.8\linewidth]{J6_omega0.eps}\vspace{0.8cm}
\includegraphics[width=0.8\linewidth]{J10_omega0_1.eps}\vspace{0.8cm}
\noindent
\caption{
The energy difference J between the triplet and the singlet 
ground states of the (1,1), (2+1,2+1), and (4+1,4+1) 
quantum Hall droplets from URHF-CI
as a function of cyclotron energy $\omega_c$.
}
\label{J4}
\end{center}\vspace{0.5cm}
\end{figure}

Correlations are included via configuration interaction method. 
Denoting the creation (annihilation) 
operators for URHF quasi-particles by $c^\dagger_{i}$ ($c_{i}$) with
the index $i$ representing the combined spin-orbit quantum numbers,
the many body Hamiltonian of the interacting system 
can be written as:
\begin{equation}
H=\sum_{ij} 
\langle i | T | j \rangle 
c^\dagger_{i} c_{j} +
\frac{1}{2}\sum_{ijkl}
V_{ijkl}
c^\dagger_{i} c^\dagger_{j} 
c_{k} c_{l} ,
\label{multiparticle}
\end{equation}       
where 
$
\langle i | T | j \rangle = \epsilon_i S_{ij} -
\langle i | V_H + V_X | j \rangle,
$
$V_{ijkl}$ are the Coulomb matrix elements in the URHF basis,
$\epsilon_i$ are the URHF eigenenegies, $V_H$ and $V_X$ are the 
Hartree and exchange operators, and 
$S_{ij}\equiv\langle\varphi_{i}|\varphi_{j}\rangle$
are the spin up and spin down orbitals overlap matrix elements.
The Hamiltonian matrix is constructed in the basis of configurations, and
diagonalized using conjugated gradient methods for different total $S_z$.
The convergence of CI calculation for the (5,5) 
droplets has been checked by increasing
the URHF basis up to $N_{S}=20$,
associated with $240~374~016$ configurations.
Fig. \ref{URHF_10e} shows the ground state energy for $N=10$ electrons
as a function of magnetic field obtained using URHF and URHF-CI methods.
The inclusion of correlations lowers total energy by $\approx 0.4Ry^*$
at $\omega_c =0$ to $0.3 Ry^*$ at $\omega_c = 0.9\omega_0$.
The results of calculated exchange interaction 
$J\equiv E_{\rm triplet} - E_{\rm singlet}$ of the half filled shells
(1,1), (3,3), (5,5)-droplets 
in magnetic field are shown in Fig. \ref{J4}. 
The spin singlet-triplet transition at $\omega_c = 0.8 \omega_0$ 
in (5,5)-molecule is equivalent to the magnetic field induced
singlet-triplet transition in a two-electron double dot
\cite{burkard-loss,HuDasSarma,Wojtek}.
The spin singlet-triplet transition at $\omega_c \approx 0.4 \omega_0$ 
is associated
with degeneracy of the LLL edge  and 2LL center orbitals.
The high magnetic field part of $J$ is quantitatively similar regardless of
the number of electrons occupying quantum dots.
In small magnetic field single particle states show complex 
structure which involves Landau level crossings and anti-crossings.
It turns out that the dependence of $J$ on the magnetic field is affected
by the single particle states of valence and core electrons,
which explains the strong dependence of $J$ on the electron numbers
in low magnetic fields.

\section{Conclusions}
We present here a microscopic theory of  
laterally coupled quantum Hall droplets with
electron numbers (N1,N2) at filling factor $\nu=2$.
Using unrestricted Hartree-Fock orbitals as the basis in
the configuration-interaction calculation we have shown that
these strongly coupled quantum dots  behave
effectively as the two-level molecule. 
When the two-level molecule is populated 
with two electrons, singlet-triplet 
spin transition can be induced by the application of high magnetic field.
The dependence on the magnetic field and the number of
core electrons of the singlet-triplet gap $J$ has been calculated.
It was shown that in ($N1\geq 5$, $N2\geq 5$) molecule 
the valence electrons can be transferred from edge to center states 
of individual dots. This might offer a possibility of performing
quantum operations on valence electrons in edge states and storage
in center orbitals.

\section{Acknowledgement} 
M.A. and P.H. acknowledge support by the NRC High 
Performance Computing project. P.H. acknowledges support by Canadian Institute for 
Advanced Research and W. D. acknowledges support by NSERC. Authors thank
A.Sachrajda and M.Pioro-Ladriere for discussions.


\end{document}